# Gain Modulation by Graphene Plasmons in Aperiodic Lattice Lasers

S. Chakraborty,[1,*] O. P. Marshall,[1,2] T. G. Folland,[1] Y.-J. Kim,[2] A. N. Grigorenko,[2] K. S. Novoselov[2,*]

[1]School of Electrical and Electronic Engineering, University of Manchester, Manchester M13 9PL, UK.
[2]School of Physics and Astronomy, University of Manchester, Manchester M13 9PL, UK.

* s.chakraborty@manchester.ac.uk; kostya@manchester.ac.uk

**Two-dimensional graphene plasmon-based technologies will enable the development of fast, compact and inexpensive active photonic elements because, unlike plasmons in other materials, graphene plasmons can be tuned via the doping level. Such tuning is harnessed within terahertz quantum cascade lasers to reversibly alter their emission. This is achieved in two key steps: First by exciting graphene plasmons within an aperiodic lattice laser and, second, by engineering photon lifetimes, linking graphene's Fermi energy with the round-trip gain. Modal gain and hence laser spectra are highly sensitive to the doping of an integrated, electrically controllable, graphene layer. Demonstration of the integrated graphene plasmon laser principle lays the foundation for a new generation of active, programmable plasmonic metamaterials with major implications across photonics, material sciences and nanotechnology.**

Among the many intriguing properties of graphene, its plasmonic characteristics are some of the most fascinating and potentially useful [1,2]. Long-lived, tunable intrinsic graphene surface plasmons (SP) have already been demonstrated in a number of experiments [3-9], including optical modulators [10,11], providing the potential for applications [12,13]. In contrast to the noble metals that are usually used in SP devices [13,14], graphene's Fermi energy, $E_F$, and carrier concentration, $n_s$ (and therefore its conductivity and SP mode properties), can be altered, for example by electrical gating and surface doping [3,15,16]. Consequently, the behavior of graphene SP-based structures can be modified in situ, without the need for structural device changes. In particular, graphene's optical and plasmonic properties are tunable in the terahertz (THz) spectral region [3,17], giving rise to the possibility of compact electrically controllable THz optical components [18]. We incorporated graphene into a plasmonic THz laser microcavity to dynamically modulate round-trip modal gain values and therefore laser emission via $E_F$. In this way gated graphene becomes a powerful tool with which to control the fundamental properties of a laser – a tool that is potentially extremely fast and all electrical in nature, with negligible electrical power requirements.

The interaction between light and matter can be altered by manipulating the electromagnetic density-of-states (DOS) using a micro resonator [19,20]. By incorporating a photonic lattice or plasmonic structure into a laser, one can control the frequency and amplification of resonant modes and hence manipulate the properties of lasing emission [21-23]. Furthermore, by breaking the regularity of these structures it is possible to modulate the photon DOS and hence light-matter interaction at several frequencies simultaneously. This technique was used recently to develop an aperiodic distributed feedback (ADFB) cavity laser with a lattice which is in essence a computer-generated hologram [24,25]. The hologram digitally encodes the Fourier transform of a desired optical filter function (multiple reflection resonances within the gain bandwidth of the laser) enabling photonic DOS manipulation at precise filter frequencies. In real space, a typical hologram lattice contains a multitude of phase-shifts; the locations and sizes of scattering sites and defects are set such that via coherent backscattering the device enters a slow light regime. Transfer matrix method (TMM) calculations of the group delay transfer function (which is intrinsically linked to the photonic DOS) of an ADFB microcavity under the influence of gain reveals infinite-gain singularities (Fig. S4, see [26] for further details). These singularities represent the frequency and gain values at which self-oscillation occurs. The ADFB microcavity can produce coherent amplification of the cavity photons via stimulated emission processes due to the build-up of phase coherence at the singularities [20].

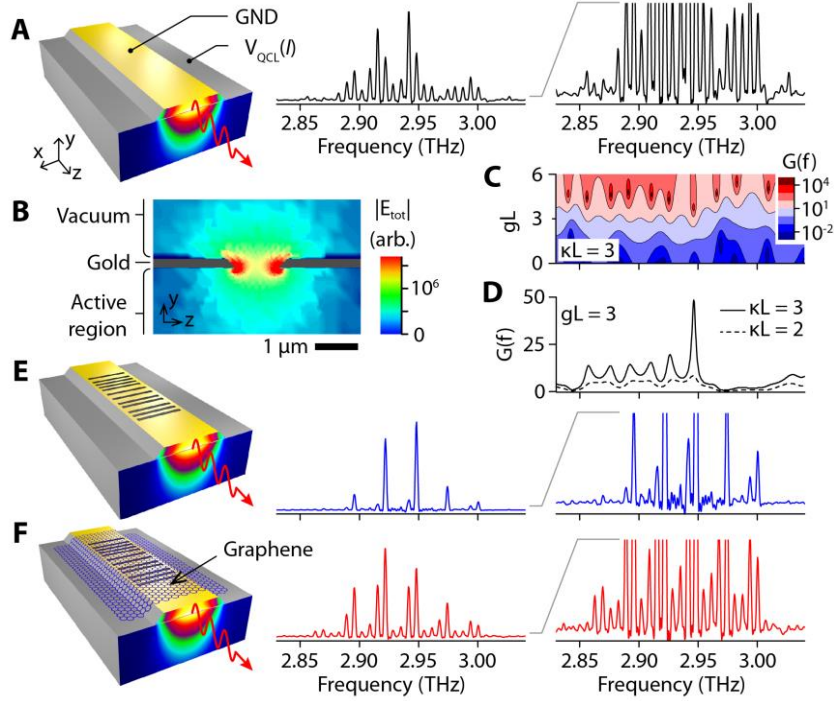

*Fig. 1. Hologram defined laser emission. (A) Schematic and typical measured emission spectra of the unperturbed Fabry-Perot QCL. (B) Simulated electric field intensity profile within a single hologram pixel (slit), f = 2.8 THz. (C) Calculated reflection gain, G(f), for a range of dimensionless material gain (gL, where L is the hologram length) values. The hologram coupling $\kappa = \Delta n/n_{eff}\Lambda$, where $n_{eff}$ is the effective refractive index, $\Delta n$ the refractive index contrast and $\Lambda$ the minimum pixel spacing. (D) G(f) for low and high $E_F$ (and $\kappa L$). Schematics and emission spectra are also shown for (E) the hologram-patterned and (F) the graphene-covered QCL.*

ADFB structures were realised in THz quantum cascade lasers (QCLs) – extremely long wavelength semiconductor lasers with active regions based on precisely engineered inter-subband transitions [27]. Such ADFB THz QCLs provide an ideal proving ground for graphene-controlled gain modulation as they employ SP-based waveguides (at a metal-semiconductor interface, Fig. 1A) [28]. The first crucial step is to excite two-dimensional (2D) plasmons in an integrated, atomically thin graphene sheet to take full leverage of the computer-generated hologram principle. Hologram pixels are introduced to the QCL waveguide as plasmonic scattering sites along the longitudinal axis of the laser ridge (Fig. 1B). By depositing an electrically gateable graphene film onto these devices our goal is to switch the THz SP at each pixel "on" or "off" by tuning $n_s$, thereby altering the photonic DOS and the degree to which the THz inter-subband gain spectra follows the hologram response. For example, by modulating the hologram pixel scattering strength we approach the DOS singularities, resulting in a dramatic increase of light-matter interaction within the QCL gain media [20]. Photon lifetimes (and hence modal gain values) are thereby enhanced, leading to selective enhancement of competing laser modes and a concomitant suppression of others.

A hologram with relatively weak feedback was chosen so that any subtle influence of graphene plasmons on laser emission was not hidden by strongly amplified photonic filtering. The hologram was designed to define multi-colour THz QCL emission [25,29] and was introduced to the metalized laser ridge surface as a series of sub-wavelength slits (Fig. 1E) [27]. At each slit, the localized removal of metal strongly influences the fundamental transverse magnetic THz eigenmode of the waveguide [25]. Finite element modelling (FEM) of the electric field across a single slit reveals strong radiative scattering of the propagating THz mode (Fig. 1B). The single-pass reflection gain (*G*) (essentially the modal gain), calculated in the frequency (*f*) and material gain (*gL*, where *L* is the hologram length) plane by using the TMM, reveals the possibility of selective mode enhancement from the microcavity resonances at reasonably achievable values of the normalised coupling factor κ*L* (Figs. 1C and 1D). This coupling is in turn dictated by the scattering strength of the hologram pixels. For further details of the FEM and TMM see [26]. Last, the key element of our design – switchable graphene plasmons – are excited in a graphene layer placed on the top of the hologram.

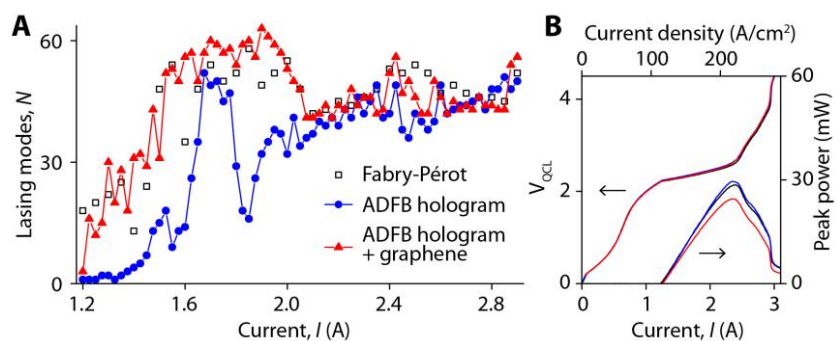

*Fig. 2. Influence of graphene deposition. (A)* Number of measured lasing modes *N* as a function of laser driving current (*I*). *(B)* Laser output power and electrical characteristics.

Four devices were fabricated and characterized, each demonstrating similar behavior, with minor differences attributable to their individual active region and hologram properties. Here we concentrate on a single representative device. Details of fabrication and testing, along with experimental results for a second device (Fig. S1), are presented in [26]. The unpatterned Fabry-Perot (FP) cavity lased on numerous longitudinal cavity modes (Fig. 1A), many of which were suppressed by implementation of the ADFB microstructure (Fig. 1E). Introduction of graphene partially 'repaired' the waveguide, reducing individual pixel scattering strengths and leading to the return of many FP modes (Fig. 1F). Laser spectra were evaluated in terms of *N*, the observed number of lasing modes (Fig. 2A), revealing the FP-like behavior of the graphene-ADFB QCL over a wide range of laser operating currents (*I*). For reference, the electrical and output power characteristics of the QCL at each stage of waveguide modification are also presented (Fig. 2B). At each stage the device displays typical THz QCL band structure alignment and misalignment features, with no appreciable changes in the absolute lasing threshold ($I_{th}$) because *g* is clamped by laser facet feedback.

In order to demonstrate electrical modification of fundamental laser gain dynamics by varying $E_F$ in the graphene, a polymer electrolyte was deposited over the device (Fig. 3A). FEM simulations of THz scattering at a single slit provides a basic understanding of the mechanisms involved (Fig. 3B). The presence of low $n_s$ (low $E_F$) graphene leads to strongly suppressed intraslit fields. Experimentally, application of gate voltage ($V_{gate}$) leads to high $n_s$ [16]. In this case, the simulated intraslit field intensities are larger. Our understanding of these results is helped by an analytical estimate of the plasmon wavelength $\lambda_{pl} = \dfrac{2\alpha E_F}{\varepsilon \hbar \omega_0} \lambda_0$, where $\alpha$ is the fine structure constant, $\omega_0$ and $\lambda_0$ are the lasing mode frequency and wavelength, respectively, and $\varepsilon$ the average permittivity surrounding the graphene (here we use $\varepsilon = 7$, the average of vacuum and GaAs) [6,8]. For $E_F = 50$ meV (typical for intrinsically doped graphene) we estimate $\lambda_{pl} \sim 1$ μm, comparable with the slit width. Consequently the electron plasma in graphene introduces a second dipole field (localised SP) within the slit, oriented opposite to the existing field, greatly reducing THz scattering (Fig. S2) [26]. On the other hand, for $E_F = 300$ meV (a reasonably achievable level by electrochemical doping) the plasmon wavelength is six times longer (large relative to the slit width) and the electron plasma moves coherently inside the slit, leading to efficient THz scattering. TMM calculations of reflection gain in the $f$-κL plane enable us to calculate the changes in modal amplification induced by raising $n_s$. Graphene-induced changes in individual pixel scattering strength (κ) can alter modal $G$ values by almost two orders of magnitude (Figs. 3C and 3D) and the group index ($n_g$) by almost one order of magnitude (Fig. S4) [26]. Owing to the reduced group velocity (slow light regime), the photon DOS is strongly enhanced around the infinite-gain singularities [20].

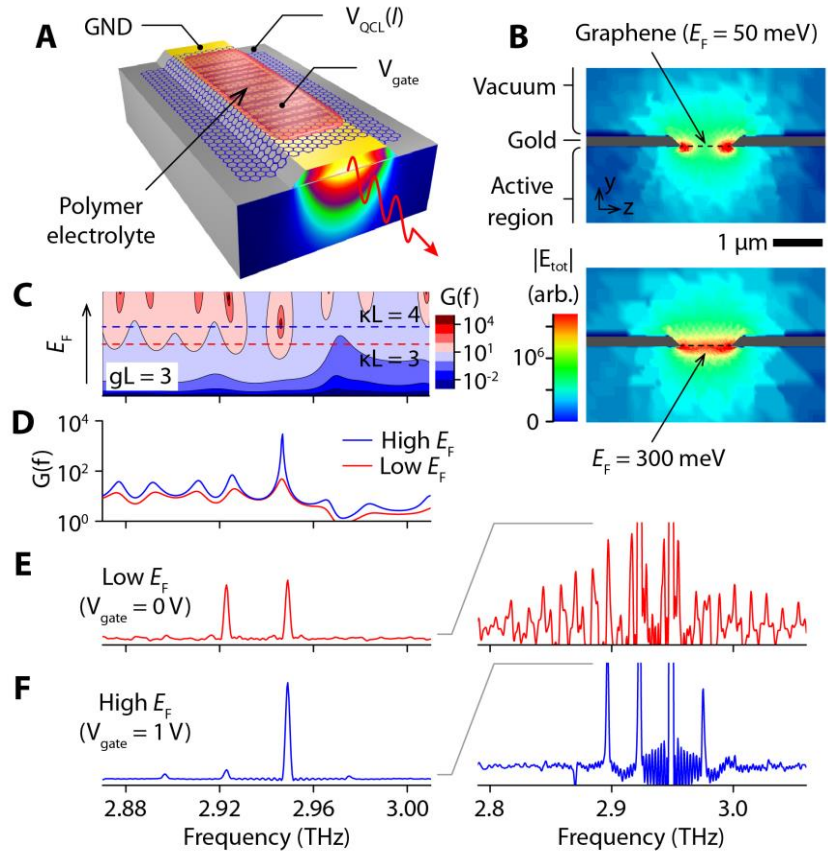

*Fig. 3. Sensitivity of laser emission to graphene doping. (A) Schematic of the polymer electrolyte-covered device. (B) Simulated electric field intensity profiles within a single hologram pixel containing low doped (upper panel) and highly doped (lower panel) graphene, f = 2.8 THz. (C) and (D) Calculated $G(f)$ as $E_F$ (and κL) is varied. Laser emission spectra measured after electrolyte deposition for (E) ungated (low $n_s$, low $E_F$) and (F) gated (high $n_s$, high $E_F$) graphene, collected just above laser threshold.*

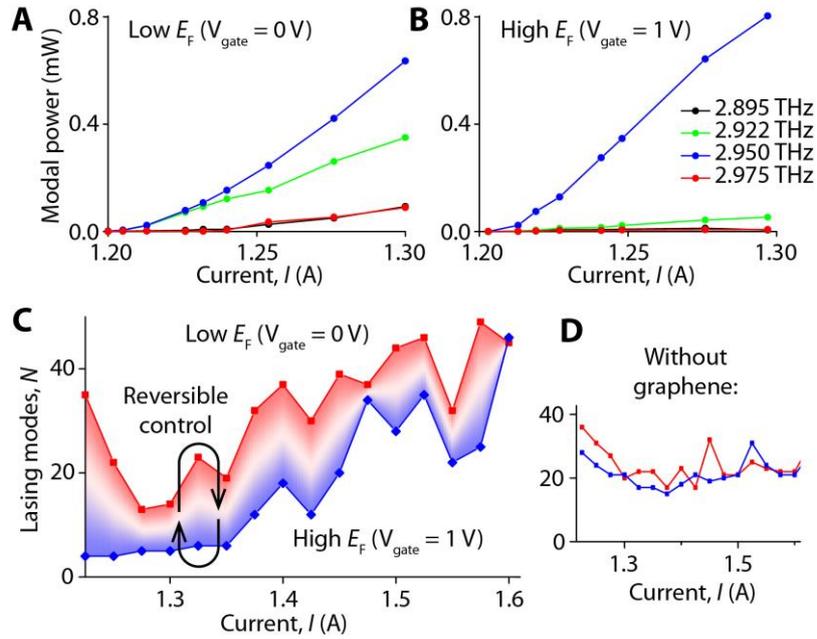

*Fig. 4. Graphene-controlled modal gain modulation.* Light-current behavior of four dominant emission frequencies for (**A**) ungated and (**B**) gated graphene. (**C**) Reversible spectral filtering (variation in N) is achieved via $V_{gate}$. (**D**) Results for electrolyte-covered QCL without graphene.

This effect has important experimental consequences (Figs. 3E and 3F). At low $E_F$, the DOS does not offer a dominant channel for inter-subband emission, and a large fraction of the emission is channelled into the FP like lasing modes. Laser emission just above $I_{th}$ is modified when we apply $V_{gate}$. By increasing $E_F$ by almost an order of magnitude, many of the FP-like lasing modes (seen at $V_{gate}$ = 0 V) are inhibited and inter-subband emission is predominantly channelled into singularities. Therefore, the high-$E_F$ graphene plasmons therefore force laser emission to be governed by the hologram response, with pure single-mode emission within each resonance band. Such a redistribution of spectral power is further observed experimentally in the light-current behavior of the four dominant modes near $I_{th}$ (Figs. 4A and 4B); with $V_{gate}$ = 1V applied we observe a strongly favoured (highest $n_g$) mode. The most dramatic reversible changes in N also occur close to $I_{th}$ (Fig. 4C) but remained appreciable over a wide current range. In contrast, when electrolyte was applied without graphene, N was insensitive to $V_{gate}$ (Fig. 4D). In this specific case, the resulting macro-scale optoelectronic functionality (close to $I_{th}$) is graphene-controlled switching between dual and single mode operation. The switching behavior is reversible up to a small finite hysteresis, as is typically observed when graphene devices are gated by solid electrolyte [16]. Demonstration of reversible graphene control is the key result of this work (not single mode lasing, which is achievable by a number of techniques). This use of graphene to define and control the fundamental gain dynamics of a laser is what sets this work apart from previously reported optical filtering in passive graphene waveguides [10,11]. Time domain modelling (TDM) provides further insight into the spatial-temporal interplay between light-field and population inversion in ADFB lasers, revealing the localization caused by the underlying aperiodicity within the hologram. Furthermore, it reveals significant changes in the inhomogeneity of the population inversion profile within the microcavity as κ is varied (Fig. S6) [26]. Any change in this profile has implications for the gain dynamics of the laser, altering the competition between lasing modes. Last, a correlation between N and pixel scattering is also seen in the TDM, indicating a direct link between the graphene-controlled electromagnetic DOS and the modal-gain of the laser (Fig. S5) [26]. We stress that the possibility to effectively control the operation of a semiconductor microcavity laser by graphene ultimately stems from unique properties of 2D graphene plasmons which allow unprecedented wavelength compression (by a factor of ~30) at small gating voltage and hence excitation of localised SP modes within the hologram pixels.

The use of electrically controllable graphene plasmons to modify active photonic systems offers a number of interesting device possibilities. In principle, each pixel (or small group of pixels) in an ADFB hologram could be independently gated, allowing individual tailoring of scattering strengths. Combined with the highly flexible multiband digital hologram approach, this would allow an operator to electronically rewrite the spectral response of a laser on demand. Furthermore, whereas programmable graphene plasmonic structures are particularly appealing for incorporation into THz lasers where spectral control is traditionally difficult, they can also be scaled to shorter-wavelength optoelectronic systems, greatly expanding their potential technological impact.

# Supplementary Materials

**Materials and Methods**

QCLs were processed from a molecular beam epitaxially (MBE) grown GaAs/Al$_{0.15}$Ga$_{0.85}$As heterostructure displaying gain tuning with alignment bias. Fabry-Perot laser cavities measured 6 mm long and 180 µm wide, with a laser ridge height of ~12 µm. Ridges were topped with a Ti/Au (20/200 nm) overlayer. ADFB holograms were introduced by focused ion beam milling (FEI Nova Nanolab 600) with gallium ions (30 kV, 1 nA, nominal spot size 50 nm). The slits run perpendicular to the laser cavity axis and measured 100 µm wide, < 1 µm long and < 1 µm deep. The hologram pattern follows that given in reference 25.

Large area, high quality graphene films (up to 99% monolayer by area) were grown by chemical vapor deposition (CVD) on 25 µm thick Cu foil. After etching of the Cu foil the graphene was transferred onto the ADFB QCLs using a polymer (PMMA) supporting layer, which was subsequently removed in acetone. The graphene serves both as part of the injection electrode and of the modulated QCL waveguide. The polymer electrolyte (LiClO$_4$, PEO) was manually deposited over the graphene-coated lasers and Au bond wires inserted for electrical biasing.

Device characterization was performed at temperatures < 10 K in a Janis ST-100 continuous flow liquid helium cryostat. Spectra were collected in pulsed laser operation (1 µs pulse width, 10 kHz repetition rate) using a Bruker Vertex 80 FTIR spectrometer (2.2 GHz resolution). THz power was measured using a calibrated thermopile (3 × 3 mm) placed directly in front of one laser facet. In later measurements $V_{gate}$ was introduced prior to device cooling; thermal cycling to room temperature was required to alter $V_{gate}$ (and $E_F$).

**Experimental Results**

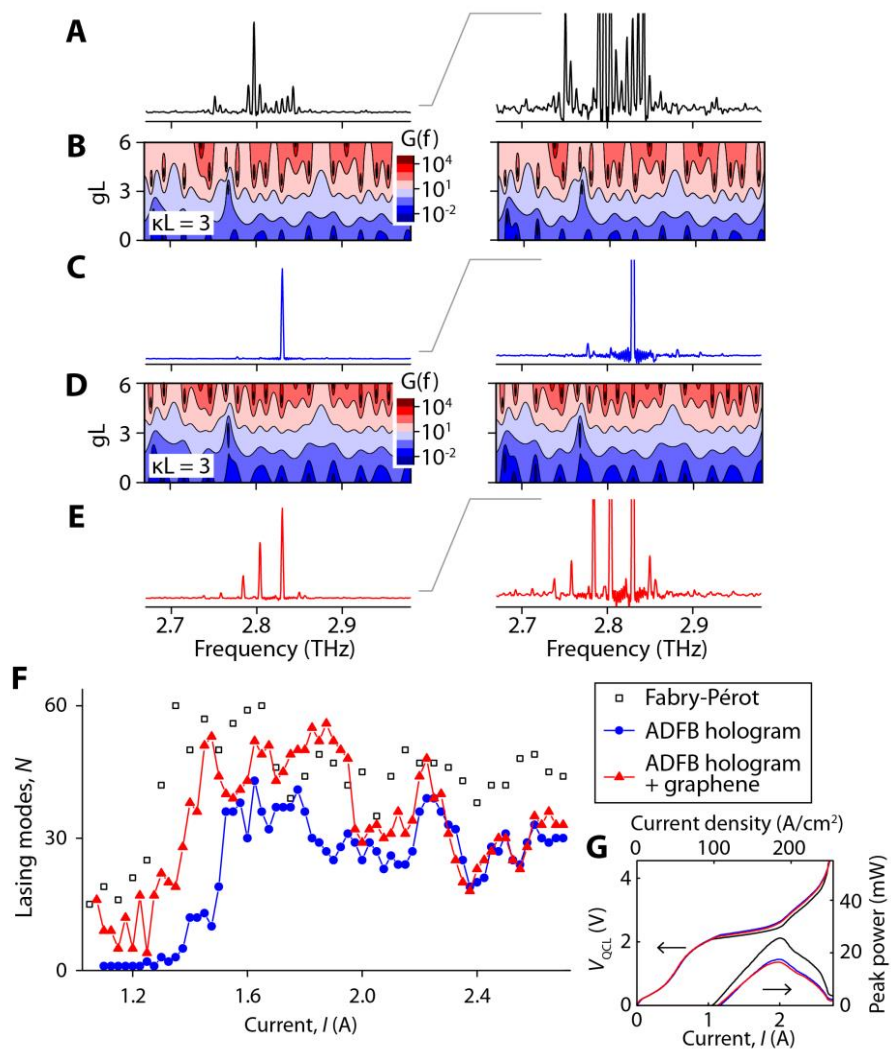

*Fig. S1. Hologram defined laser emission in a second device. (A) Typical measured emission spectra of a second unperturbed FP QCL. (B) Reflection gain G(f) and (C) typical measured emission spectra after ADFB hologram patterning. (D) Calculated G(f) and (E) measured emission spectrum after introduction of graphene. (F) Number of observed lasing modes at each stage of laser modification. (G) Laser output power and electrical characteristics.*

## Passive graphene-modified ADFB waveguide modelling

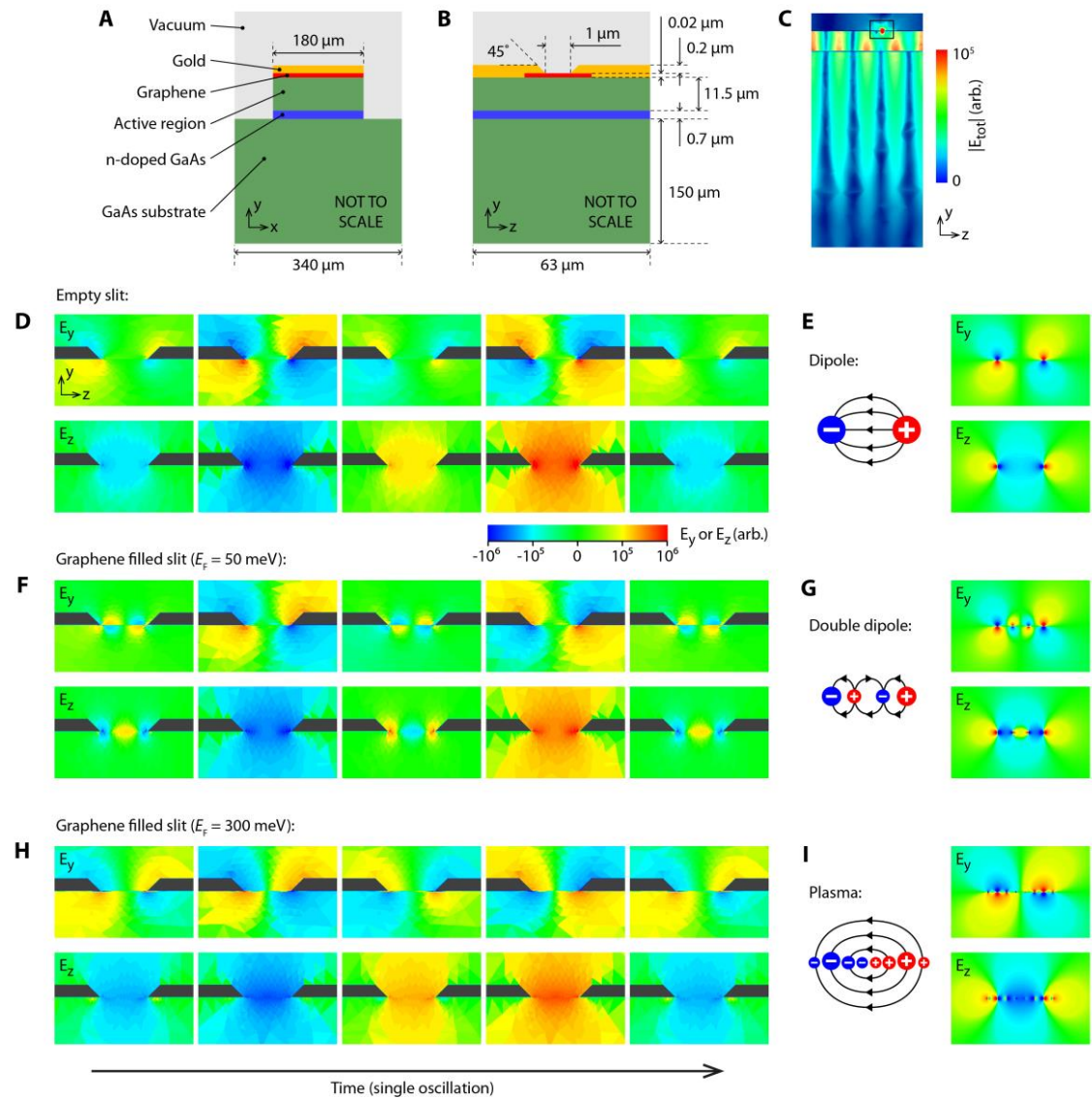

***Fig. S2. TEM of a single slit.*** *(A) and (B) Schematics of the modelled structure. (C) Longitudinal section through the simulated electric field profile in the waveguide section. (D) Time-varying electric field components in the region of the slit, which closely resemble (E) the field components of a simple dipole charge. (F) A slit containing low doped graphene resembles (G) double, counter-oriented dipole charges. (H) A slit containing highly doped graphene resembles (I) distributed charges.*

In this work the ADFB hologram was patterned as slits in the Ti/Au uppermost layer of the QCL waveguide, locally altering the optical mode profile and propagation index to cause reflections at defined locations (i.e. distributed feedback). The waveguide regions without Au are extremely short (< 1 µm) compared with the radiation wavelength in the material (~30 µm). Consequently, basic 2D modelling approaches only indicate that introduction of graphene perturbs the THz mode, but do not allow us to specify the exact interaction mechanism. In a more rigorous study of the 3D field behavior, commercially available FEM software (HFSS) was employed. Due to the increased computational complexity it was not possible to solve for the 3D eigenmodes of the entire structure. Instead, THz wave ports were introduced to each end of a short modelled section of QCL waveguide containing a single slit (Figs. S2A and S2B). This simple structure provides insight into the mode perturbation induced by a single hologram element. Results for a single slit can then be extrapolated to explain the behavior of a full ADFB QCL. Simulations were run over a range of frequencies (2.7 to 3.1 THz) and a Drude-Lorentz model was used to calculate the material optical properties at each frequency. Periodic boundary conditions ensured standing wave solutions. Thin film planar graphene was introduced to the slit, undercutting the Au slightly to avoid graphene edge effects. Furthermore, graphene was defined with anisotropic properties (conductive in the plane of the sheet but equivalent to the underlying active region perpendicular to the sheet) to mitigate any inaccuracies stemming from the thin film approach. A longitudinal section was taken through the resulting profile (see Figure 1B). Following classical electrodynamics we can treat such sub-wavelength apertures as simple dipole scattering sites, their dipole moment deduced from the electric field intensity across the aperture surface [29]. Large intra-slit electric fields (due to dipole charges produced by the metallic slit edges) therefore lead to strong radiative scattering of the propagating THz mode. The electric field magnitude profile (without graphene and at a frequency of 2.8 THz) also contains a series of maxima along the laser axis due to the standing-wave THz radiation, along with intense localised fields in the vicinity of the slit (Fig. S2C). The time varying field components for an empty slit behave like that of a simple oscillating dipole charge distribution (Figs. S2D-E). For slits containing graphene doped at $E_F$ = 50 and 300 meV the fields resemble those of a double dipole and distributed charges respectively (Figs. S2F-I). Note that the precise position and geometry of the slit influence the magnitude of the field profile results, but not their form. An in-depth study of this geometric dependence is beyond the scope of this work. Note that the tapered slit edges smooth the transition between suppressed and enhanced scattering (for low and high $E_F$ respectively). Finally, it must also be pointed out that experimental verification these modelling results, by direct probing of the graphene plasmon, is not possible in the existing devices due to the presence of the relatively thick polymer electrolyte. Any future direct measurements of this nature would require modification of the device architecture.

**Active graphene-modified active ADFB laser modelling - Transfer Matrix Method (TMM)**

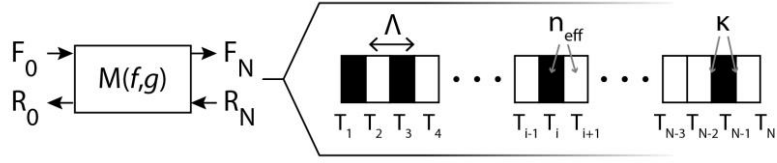

***Fig. S3. Transfer Matrix Method (TMM) approach.*** *The transfer matrix M consists of a series of sub-matrices which follow the ADFB hologram design.*

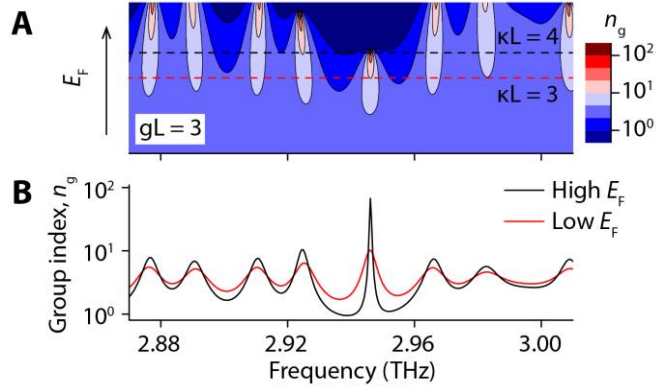

***Fig. S4. Graphene-controlled group index.*** *(A) and (B) Group index ($n_g$) as $E_F$ (and therefore $\kappa L$) is varied, calculated using the TMM approach. In the QCLs reported here, due to the presence of facet reflections we estimate $gL \sim 3$ (L = 2.6 mm, the hologram length).*

To calculate the properties of the ADFB hologram in the presence of gain we exploit a TMM based on that in refs [30]. The hologram consists of an arbitrary arrangement of high and low refractive index layers of width $\Lambda/2$, where $\Lambda$ is the period of a uniform grating. The forward (*F*) and backward (*R*) propagating waves in the structure can be related through a linear transfer matrix (*M*). This matrix can be expressed as the product of a series of sub-matrices ($T_i$), each governing the propagation and scattering within each of these elements (Fig. S3). For simplicity, propagation was performed in each element according to a spectrally flat effective refractive index ($n_{eff}$), which possesses a complex component to represent material gain (*g*). Scattering between different layers was calculated from a refractive index step ($\Delta n$), defined by the dimensionless coupling constant ($\kappa L$), where *L* is the hologram length, $\Delta n = \kappa \lambda_B / 2$ and $\lambda_B = 2 n_{eff} \Lambda$. In principle the refractive index step can be real or imaginary (in the simulations presented this work it is assumed to be real). For all simulations $\Lambda$ = 12.8 μm and $n_{eff}$ = 3.6757, appropriate for the THz QCL presented in the main text. We can express *M* in terms of its coefficients $T_{ij}$;

$$\begin{pmatrix} F_0 \\ R_0 \end{pmatrix} = M \begin{pmatrix} F_N \\ R_N \end{pmatrix} = \prod_{i=1}^{N} T_i \begin{pmatrix} F_N \\ R_N \end{pmatrix} = \begin{pmatrix} T_{11} & T_{12} \\ T_{21} & T_{22} \end{pmatrix} \begin{pmatrix} F_N \\ R_N \end{pmatrix}$$

from which the reflection and transmission gain coefficients can be defined; $r = T_{12}/T_{11}$ and $t = 1/T_{11}$ (and corresponding power reflection and transmission: $R = |r|^2$, $T = |t|^2$). Furthermore, using the phase ($\varphi_t$) of the complex transmission gain coefficient we can also calculate the spectral group delay, and hence group index of waves propagating in the hologram using the following expression:

$$\tau_g(\omega) = \frac{\partial \varphi_t}{\partial \omega}$$

We calculate the reflection gain (and the group index) transfer functions over the full (*f,g*) and (*f,κL*) planes, for fixed *κL* and *g* respectively. These results can be represented by a contour plot, where the reflectivity gain (or the group index) is represented by contour information as a variable of *f* and *g* (or *κL*) (Fig. S4). The condition for self-oscillation in the structures is provided by the singularity points, where *R* tends to infinity, corresponding to a finite output for zero input (Fig. 1C).

**Active graphene-modified active ADFB laser modelling - Time Domain Modelling (TDM)**

TDM is based upon the coupled wave equations from DFB theory and was originally developed to simulate (periodic) DFB semiconductor laser behavior [31]. In this work the TDM code described in reference 31 was modified to analyze the response of the hologram embedded within a laser cavity with or without FP facets. The modified TDM code is a powerful tool incorporating all aspects of the laser, from the gain (as dictated by the full laser rate equations) to the ADFB hologram structure. As such, whilst the TMM allows us to calculate the singularities of a given structure, the TDM provides insight into the interaction between hologram and gain medium. Put simply, the TDM code works by splitting a laser cavity into multiple discrete sections and converting the coupled wave equations to propagation matrices through these sections. The ADFB hologram was implemented within the TDM code by incorporating multiple defects, introduced as phase shifts of varying magnitude [29]. Starting with the random process of spontaneous emission and iteratively solving the coupled propagation matrices through space and time, it is possible to simulate both the output spectrum and internal electron/photon populations of a laser. This allows us to probe how controlling the hologram strength using graphene plasmons (represented by the normalized coupling factor $\kappa L$) influences the spatial gain dynamics and spectrum.

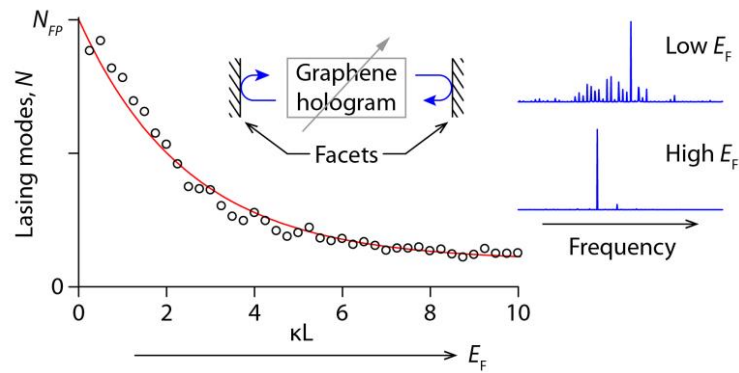

*Fig. S5. Time Domain Modelling (TDM).* The number of lasing modes N (circles) falls exponentially with increasing $\kappa L$ (concomitant with increasing $E_F$). When $\kappa L = 0$ lasing is achieved on the maximum number of FP cavity modes ($N_{FP}$). Left inset: Schematic of modelled laser waveguide. Right insets: Selected calculated laser emission spectra.

First we study the emission spectra produced by the compound system that consists of a hologram embedded into a FP laser cavity, matching the experimentally demonstrated structure. For ease of quantitative comparison between numerical and experimental results the number of lasing modes (*N*) was chosen as the clearest metric of ADFB-modified laser emission. Lasing modes were defined as any spectral peak lying above the statistical noise floor. Multiple simulations were run for each $\kappa L$ value, improving the statistical significance of the averaged *N*. As expected the results show that ADFB laser emission is highly sensitive to scattering strength, with *N* exponentially decreasing as $\kappa L$ is increased from zero (equivalent to an unperturbed FP cavity) (Fig. S5). It is this fundamental control of the lasing process via dynamic modulation of the scattering strength that is accessed via graphene doping. For insight into how this modulation influences the gain dynamics and characteristics of graphene controlled lasing we now study the spatial fluctuation of population inversion within the hologram lattice (Fig. S6). Laser facets were removed from this simulation to aid in clarity. At low scattering strengths there is insufficient feedback for lasing, and as a result there is uniform spatial coupling between the electric field and electron distribution. For moderate coupling strengths the coherent backscattering process begins, and the electron concentration is suppressed at the edges of the grating. This is a consequence of the diffusive nature of light in weakly scattering media, leading to high photon density at the edges of the structure. When the feedback becomes strong the lasing mode becomes spatially modulated in the propagation direction. The electron concentration is significantly reduced at specific 'hot spots' within the lattice. In the vicinity of the infinite-gain singularity points there is an enhanced stimulated emission rate and reduced electron population. Overall, TDM simulations suggest a high degree of localization due to the underlying aperiodicity within the lattice. Eventually localization becomes so strong that dynamic mode competition sets in between the modes originating from the singularities [20]. It is this competition which leads to the mode switching observed in the insets of Figure S5.

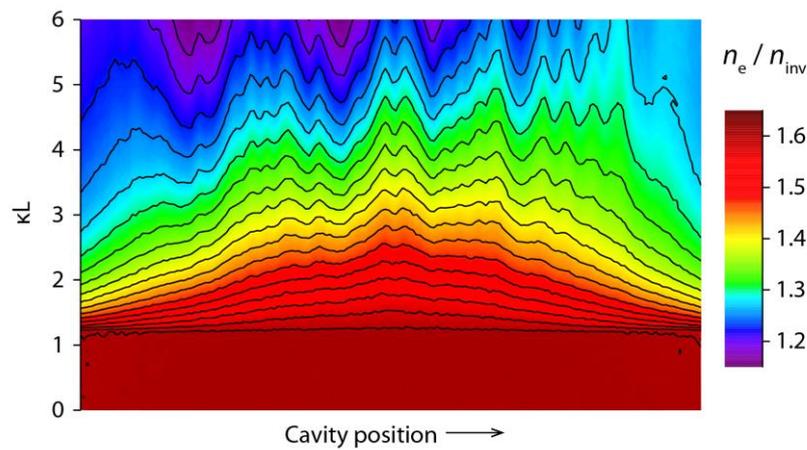

**Fig. S6. Inhomogeneous population inversion.** The electron concentration ($n_e$) profile (normalised to the inversion concentration, $n_{inv}$) within the ADFB microcavity for a range of $\kappa L$.